\documentclass{article}%
\usepackage{amsfonts}
\usepackage{amsmath}
\usepackage{amssymb}
\usepackage{graphicx}%
\newtheorem{theorem}{Theorem}

\newtheorem{definition}[theorem]{Definition}

\newtheorem{lemma}[theorem]{Lemma}

\newtheorem{proposition}[theorem]{Proposition}

\newenvironment{proof}[1][Proof]{\noindent\textbf{#1.} }{\ \rule{0.5em}{0.5em}}
\begin{document}

\title{Presymplectic representation of bi-Hamiltonian chains}
\author{Maciej B\l aszak\thanks{Supported partially by KBN research grant No. 1 P03B
111 27}\\Institute of Physics, A. Mickiewicz University, \\61-614 Pozna\'{n}, Umultowska 85}
\maketitle

\begin{abstract}
Liouville integrable systems, which have bi-Hamiltonian representation of the
Gel'fand-Zakharevich type, are considered. Bi-presymplectic representation of
one-Casimir bi-Hamiltonian chains and weakly bi-presymplectic representation
of multi-Casimir bi-Hamiltonian chains are constructed. The reduction
procedure for Poisson and presymplectic structures is presented.

\end{abstract}

\section{Introduction}

The bi-Poisson formulation of finite dimensional integrable Hamiltonian
systems has been systematically developed for the last two decades (see
\cite{b1} and the literature quoted there). It has been found that most of the
known Liouville integrable finite dimensional systems have more then one
Hamiltonian representation. Moreover, in the majority of known cases, both
Poisson structures of a given flow are degenerated. Perhaps this is the reason
why such an important property of integrable systems was discovered so late,
relative to the age of classical mechanics. For such systems, related
bi-Poisson (bi-Hamiltonian) commuting vector fields belong to one or more
bi-Hamiltonian chains starting and terminating with Casimirs of respective
Poisson structures. An important aspect of such a construction is its relation
to the recently developed geometric separability theory \cite{1}-\cite{m3}.
Actually, the necessary condition for the existence of separation coordinates
is the reducibility of one of the Poisson structures onto a symplectic leaf of
the other one. An important fact is, that the whole procedure of variables
separation is almost algoritmic.

On the other hand, it is well known from the classical mechanics, that if the
Poisson structure is nondegenerate, i.e. if the rank of the Poisson tensor is
equal to the dimension of a phase space, then the phase space becomes a
symplectic manifold with a symplectic structure being just the inverse of the
Poisson structure. In such a case there exists an alternative (dual)
description of Hamiltonian vector fields in the language of symplectic
geometry. So, a natural question arises, whether one can construct such a dual
picture in the degenerated case, when there is no natural inverse of the
Poisson tensor \cite{dub}.

A positive answer to this question is presented in next Sections of the paper.
A dual presymplectic picture will be constructed for bi-Hamiltonian chains
with one Casimir as well as with many Casimirs. The paper is organized as
follows. In this Section we recall some elementary facts from the Poisson and
presymplectic geometry. In Section 2 we introduce notions of dual pairs,
compatible pairs and Poisson pairs and investigate some of their properties.
In Section 3, applying the results of the previous Section, we construct a
presymplectic representation of Poisson chains. In Section 4 the deformation
reduction procedure for Poisson and presymplectic chains is presented. Such a
reduction is crucial for separability of underlying dynamical systems.
Finally, in Section 5, we illustrate the presented theory by a nontrivial example.

Given a manifold $\mathcal{M}$ of $\dim\mathcal{M}=m,$ a \emph{Poisson
operator} $\Pi$ of corank $r$ on $\mathcal{M}$ is a bivector $\Pi\in
\Lambda^{2}(\mathcal{M})$ with vanishing Schouten bracket:
\begin{equation}
\lbrack\Pi,\Pi]_{S}=0, \label{Schouten}%
\end{equation}
whose kernel is spanned by exact one-forms
\[
\ker\Pi=Sp\{dc_{i}\}_{i=1,...,r}.
\]
The symbol $d$ denotes the operator of exterior derivative.\ In a given
coordinate system $(x^{1},\ldots,x^{m})$ on $\mathcal{M}$ we have
\[
\Pi=\sum\limits_{i<j}^{m}\Pi^{ij}\frac{\partial}{\partial x^{i}}\wedge
\frac{\partial}{\partial x^{j}},
\]
while the Poisson property (\ref{Schouten}) takes the form
\[
\sum_{l}(\Pi^{lj}\partial_{l}\Pi^{ik}+\Pi^{il}\partial_{l}\Pi^{kj}+\Pi
^{kl}\partial_{l}\Pi^{ji})=0,\ \ \ \partial_{i}:=\frac{\partial}{\partial
x^{i}}.
\]
\ A function $c:\mathcal{M}\rightarrow\mathbb{R}$ is called the \emph{Casimir
function} of the Poisson operator $\Pi$ if $\Pi dc=0$. A linear combination
$\Pi_{\lambda}=\Pi_{1}-\lambda\Pi_{0}$ ($\lambda\in\mathbb{R}$) of two Poisson
operators $\Pi_{0}$ and $\Pi_{1}$ is called a \emph{Poisson pencil} if the
operator $\Pi_{\lambda}$ is Poissonian for any value of the parameter
$\lambda$. In this case we say that $\Pi_{0}$ and $\Pi_{1}$ are
\emph{compatible}$.$ A vector field $X_{F}$ related to a function $F$ through
the relation
\begin{equation}
X_{F}=\Pi dF \label{1.1}%
\end{equation}
is called a Hamiltonian vector field with respect to the Poisson operator
$\Pi$. It is also important to note that if $X$ is any vector field on
$\mathcal{M}$ that is Hamiltonian with respect to $\Pi,$ then $L_{X}\Pi=0$
,where $L_{X}$ is the Lie-derivative operator in the direction $X$.

Further, a \emph{presymplectic operator} $\Omega$ on $\mathcal{M}$ defines a
2-form that is closed, i.e. $d\Omega=0,$ degenerated in general. In the
coordinate system $(x^{1},\ldots,x^{m})$ on $\mathcal{M}$ we can always
represent $\Omega$ as
\[
\Omega=\sum\limits_{i<j}^{m}\Omega_{ij}dx^{i}\wedge dx^{j},
\]
where the closeness condition takes the form
\[
\partial_{i}\Omega_{jk}+\partial_{k}\Omega_{ij}+\partial_{j}\Omega_{ki}=0.
\]
Moreover, the kernel of any presymplectic form is always an integrable
distribution. A vector field $X^{F}$ related to a function $F$ by the relation%
\begin{equation}
\Omega X^{F}=dF \label{1.2}%
\end{equation}
is called the inverse Hamiltonian vector field with respect to the
presymplectic operator $\Omega$. Generally, if $\Omega$ is a closed two-form
and $X$ is an arbitrary vector field then
\begin{equation}
L_{X}\Omega=d(\Omega X). \label{1.2a}%
\end{equation}
Hence, if $\Omega(Y)=0$ for some vector field $Y$ on $\mathcal{M}$ then
$L_{Y}\Omega=0$. Notice that contrary to the Poisson case, a linear
combination of two presymplectic operators is always presymplectic.

Poisson tensor $\Pi,$ considered as the mapping $\Pi:T^{\ast}\mathcal{M}%
\rightarrow T\mathcal{M}$, induces a Lie bracket on the space $C^{\infty
}(\mathcal{M})$ of all smooth real-valued functions on $\mathcal{M}$
\begin{equation}
\left\{  .,.\right\}  _{\Pi}:C^{\infty}(\mathcal{M})\times C^{\infty
}(\mathcal{M})\rightarrow C^{\infty}(\mathcal{M})\text{, \ }\left\{
F,G\right\}  _{\Pi}\overset{\mathrm{def}}{=}\left\langle dF,\Pi
\,dG\right\rangle =\Pi(dF,dG), \label{bracket}%
\end{equation}
(where $\left\langle .,.\right\rangle $ is the dual map between $T\mathcal{M}$
and $T^{\ast}\mathcal{M}$) which is skew-symmetric and satisfies Jacobi
identity. It is called a \emph{Poisson bracket}.

When a Poisson operator $\Pi$ is nondegenerate, one can always define its
inverse $\Omega=\Pi^{-1},$ called a \emph{symplectic operator}$,$ and then
equations (\ref{1.1}) and (\ref{1.2}) are equivalent. Moreover, any
Hamiltonian vector field with respect to $\Pi$ is simultaneously the inverse
Hamiltonian with respect to $\Omega$ and $X_{F}=X^{F}.$ Finally, symplectic
operator $\Omega$ defines the same Poisson bracket as the related Poisson
operator $\Pi$
\begin{equation}
\{F,G\}^{\Omega}:=\Omega(X^{F},X^{G})=<\Omega X^{F},X^{G}>=<dF,\Pi
dG>=\{F,G\}_{\Pi}.
\end{equation}
The equivalence is destroyed in the case of degeneracy. First, one cannot
define $\Omega$ as the inverse of $\Pi.$ Second, for degenerated $\Pi$
equation (\ref{1.1}) is valid for an arbitrary function $F$ (as in the
nondegenerate case), while for degenerated $\Omega$ and an arbitrary $F$ there
is no such vector field $X^{F}$ that (\ref{1.2}) is fulfilled. It means that
equation (\ref{1.2}) is valid only for a particular class of functions
(contrary to the nondegenerate case). Finally it is not clear how to define a
Poisson bracket with respect to a presymplectic form.

\section{Dual Poisson-presymplectic pairs and compa- tible structures}

In this Section we introduce basic objects important for the theory further
developed and we investigate some of their properties. As the concept of dual
pairs was introduced and developed for the first time in our previous paper
\cite{dirac}, here we only recall their main properties. Let us remark that
the concept of dual Poisson-presymplectic pairs \cite{dirac}, which we are
going to apply to bi-Poisson chains, is a useful particular realization of the
concept of Poisson brackets on presymplectic manifolds, presented by
B.A.Dubrovin at al. \cite{dub}.

Consider a smooth manifold \emph{M} of dimension $m$ equipped with a pair of
antisymmetric operators $\Pi$, $\Omega$.

\begin{definition}
\label{dualdef}A pair of antisymmetric tensor fields $(\Pi,\Omega)$ such that
$\Pi:T^{\ast}\mathcal{M}\rightarrow T\mathcal{M}$ , i.e. $\Pi$ is twice
contravariant, and $\Omega:T\mathcal{M}\rightarrow T^{\ast}\mathcal{M}$ , i.e.
$\Omega$ is twice covariant, is called a dual pair if there exists $r$
one-forms $\alpha_{i}$, $i=1,...,r$ and $r$ linearly independent vector fields
$Z_{i}$, $i=1,\ldots,r$ such that the following conditions are satisfied:

\begin{enumerate}
\item $\alpha_{i}(Z_{j})=\delta_{ij}$ for all $i,j=1,\ldots r$.

\item The kernel of $\Pi$ is spanned by all $\alpha_{i}$, $\ker(\Pi
)=Sp\{\alpha_{i}\}_{i=1..r}$.

\item The kernel of $\Omega$ is spanned by all the vector fields $Z_{i}$,
$\ker(\Omega)=Sp\{Z_{i}\}_{i=1..r}$.

\item The following\emph{\ partition of unity} holds on $T\mathcal{M}$%
\begin{equation}
I=\Pi\Omega+%
{\textstyle\sum\limits_{i=1}^{r}}
Z_{i}\otimes\alpha_{i} \label{partition}%
\end{equation}
where $\otimes$ denotes the tensor product.
\end{enumerate}
\end{definition}

Notice that the partition of unity (\ref{partition}) on $T^{\ast}\mathcal{M}$
takes the form
\begin{equation}
I=\Omega\Pi+%
{\textstyle\sum\limits_{i=1}^{r}}
\alpha_{i}\otimes Z_{i}.\label{2.1a}%
\end{equation}
Let us choose the basic one-forms $\alpha_{i}$ in such a way that $\alpha
_{i}=dc_{i}$ and let us denote a foliation of $\mathcal{M~}$\ given by the
functions $c_{i}$ by $\mathcal{N}$. This foliation consists of the leaves
$\mathcal{N}_{\nu}=\{x\in M:c_{i}(x)=\nu_{i}$, $i=1,\ldots,r\}$, $\nu=(\nu
_{r},\ldots,\nu_{r})$. Condition 1 of the above definition implies that the
distribution $\mathcal{Z}$ spanned by the vector fields $Z_{i}$ is transversal
to the foliation $\mathcal{N}$. Thus, for any $x\in\mathcal{M}$ we have%
\begin{equation}
T_{x}\mathcal{M}=T_{x}\mathcal{N}_{\nu}\oplus\mathcal{Z}_{x}\text{, \ \ }%
T_{x}^{\ast}\mathcal{M}=T_{x}^{\ast}\mathcal{N}_{\nu}\oplus\mathcal{Z}%
_{x}^{\ast}\label{2.1b}%
\end{equation}
where $\mathcal{N}_{\nu}$ is a leaf from the foliation $\mathcal{N}$ that
passes through $x$, the symbol $\oplus$ denotes the direct sum of the vector
spaces, $\mathcal{Z}_{x}$ is the subspace of $T_{x}\mathcal{M}$ spanned by the
vectors $Z_{i}$ at this point, $T_{x}^{\ast}\mathcal{N}_{\nu}$ is the
annihilator of $\mathcal{Z}_{x}$ and $\mathcal{Z}_{x}^{\ast}$ is the
annihilator of $T_{x}\mathcal{N}_{\nu}$. Condition 2 of the above definition
implies that $\operatorname{Im}(\Pi)$ $=T\mathcal{N}$, Condition 3 means that
$\operatorname{Im}(\Omega)$ $=T^{\ast}\mathcal{N}$ and Condition 4 describes
the degree of degeneracy of our pair.

\begin{definition}
\label{Ppdef}A dual pair $(\Pi,\Omega)$ is called a dual Poisson-presymplectic
pair (in short: dual \emph{P-p }pair) if $\Pi$ is a Poisson bivector and if
$\Omega$ is a closed 2-form.
\end{definition}

Notice that in the case when a dual P-p pair has no degeneration ($r=0$) we
get the usual Poisson-symplectic pair of mutually inverse operators, since
(\ref{partition}) reads then as $I=\Pi\Omega$. Moreover, for a degenerated
case, when $r\neq0,$ as $\Omega$ is presymplectic, then $\ker(\Omega)$ is an
integrable distribution with $[Z_{i},Z_{j}]=0,\ i,j=1,...,r$ and for $\Pi$
Poisson, $\alpha_{i}$ are exact one forms generated by Casimir functions:
$\alpha_{i}=dc_{i},\ i=1,...,r.$ The commutativity of $Z_{i}$ follows from
Condition 1 of Definition \ref{dualdef}. The following Lemma will be useful in
further considerations.

\begin{lemma}
\label{l1}Let $(\Pi,\Omega)$ be a dual P-p pair, then
\[
L_{Z_{i}}\Pi=0,~i=1,...,r.
\]

\end{lemma}

Assume that $(\Pi,\Omega)$ is a dual P-p pair and
\begin{equation}
\Pi dF=X_{F}\label{2.2b}%
\end{equation}
is a Hamiltonian vector field with respect to $\Pi.$ Applying $\Omega$ to both
sides of (\ref{2.2b}) and using the decomposition (\ref{2.1a}) we get%
\begin{equation}
dF=\Omega(X_{F})+%
{\displaystyle\sum\limits_{i=1}^{r}}
Z_{i}(F)dc_{i},\label{2.2a}%
\end{equation}
which reconstructs $dF$ from $X_{F}$ and $Z_{i}(F)$ in the case of degenerated
Poisson structure $\Pi$. In that sense $\Omega$ plays the role of the
"inverse" of $\Pi.$ Notice that inverse Hamiltonian vector fields with respect
to $\Omega$ are related to functions which are annihilated by $\ker(\Omega),$
i.e. $Z_{i}(F)=0\ ,i=1,...,r.$ Then, equation (\ref{2.2a}) reduces to
(\ref{1.2}) with $\Omega(X_{F})=\Omega(X^{F})$. It means that $X_{F}$ is not
only a Hamiltonian but also inverse Hamiltonian vector field related to the
same Hamiltonian function $F$. Moreover, it is a gauge freedom for inverse
Hamiltonian vector fields $X^{F}$ with respect to $\Omega.$ Indeed, applying
$\Pi$ to both sides of equation (\ref{1.2}) and using decomposition
(\ref{partition}) one gets
\[
X^{F}-X_{F}=\sum_{i}X^{F}(c_{i})Z_{i}.
\]
It means that an inverse Hamiltonian vector field $X^{F}$ is simultaneously a
Hamiltonian vector field, i.e. $X^{F}=X_{F},$ if $X^{F}$ annihilates the
kernel of $\Pi.$

The definition of dual objects is not unique and questions about the 'gauge
freedom' can be posed. A possible realization of such a freedom is as follows:
given a dual P-p pair $(\Pi,\Omega)$ we are looking for possible deformations
of $\Omega$ to get a new presymplectic form $\Omega^{\prime}$ ensuring that
$(\Pi,\Omega^{\prime})$ is dual again. Another possibility is related to a
gauge freedom for the operator $\Pi$, i.e. how can we deform $\Pi$ to a new
Poisson bivector $\Pi^{\prime}$ so that $(\Pi^{\prime},\Omega)$ is also the
dual pair. An example of such a gauge freedom is given in the following proposition:

\begin{proposition}
\label{p1}Let $(\Pi,\Omega)$ be a dual P-p pair as in definitions
\ref{dualdef} and \ref{Ppdef}. Suppose that $F_{i}$ are real functions on
$\mathcal{M}$ related to vector fields $K_{i}$ which are simultaneously
Hamiltonian and inverse Hamiltonian with respect to $(\Pi,\Omega)$ pair
\[
dF_{i}=\Omega K_{i},\ \ \ K_{i}=\Pi dF_{i},\ \ i=1,...,r.
\]
Then:

\begin{description}
\item[(i)]
\[
\Omega^{\prime}=\Omega+%
{\textstyle\sum\limits_{i}}
dF_{i}\wedge dc_{i},
\]
is a dual to $\Pi$ presymplectic two-form, provided that
\[
\Pi(dF_{i},dF_{j})=0\text{ for all }i,j\text{. }%
\]

\item[(ii)]
\[
\Pi^{\prime}=\Pi+%
{\textstyle\sum\limits_{i}}
Z_{i}\wedge K_{i}%
\]
is a dual to $\Omega$ Poisson bivector, provided that
\[
\Omega(K_{i},K_{j})=0\ \ \text{ for all }i,j.
\]

\end{description}
\end{proposition}

Let us now turn our attention to brackets induced on the space $C^{\infty
}(\mathcal{M}).$ We know that the Poisson operator $\Pi$ turns $C^{\infty
}(\mathcal{M})$ into a Poisson algebra with the Poisson bracket (\ref{bracket}%
)
\[
\left\{  F,G\right\}  _{\Pi}=\Pi(dF,dG)=\left\langle dF,\Pi\,dG\right\rangle .
\]
In case when $\Omega$ is a part of a dual P-p pair we can define the above
bracket trough the $\Omega$ in the following way:

\begin{lemma}
\label{l3}Let $(\Pi,\Omega)$ be a dual P-p pair. Define a new bracket on
$C^{\infty}(\mathcal{M})$
\[
\left\{  F,G\right\}  ^{\Omega}:=\Omega(X_{F},X_{G})=<\Omega X_{F}%
,X_{G}>,\ \ \ \ \ X_{F}=\Pi dF.
\]
Then $\left\{  \cdot,\cdot\right\}  ^{\Omega}=\left\{  \cdot,\cdot\right\}
_{\Pi}$ i.e. both brackets are identical.
\end{lemma}

The proofs of Lemma 3, Lemma 5 and Proposition 4, as well as more details on
the concept of dual P-p pairs the reader can find in \cite{dirac}. \

Now we pass to the concept of compatibility.

\begin{definition}
\label{d4a}A Poisson bivector $\Pi$ and presymplectic two-form $\Omega$ are
called a compatible P-p pair if $\Omega_{D}:=\Omega\Pi\Omega$ is presymplectic.
\end{definition}

As well known (see for example \cite{b1}) if $(\Pi,\Omega)$ is a compatible
P-p pair, then the second order tensor $\Phi=\Pi\Omega:\ T\mathcal{M}%
\rightarrow$ $T\mathcal{M}$ has vanishing Nijenhuis torsion
\[
L_{\Phi\tau}\Phi-\Phi L_{\tau}\Phi=0,\ \ \ \ \forall\tau\in T\mathcal{M},
\]
and is called a hereditary operator or recursion operator. Moreover $\Pi
_{D}:=\Pi\Omega\Pi$ is a Poisson bivector. Observe that a dual P-p pair
$(\Pi_{0},\Omega_{0})$ is a trivial example of a compatible pair as
\begin{equation}
\Omega_{D}=\Omega_{0}\Pi_{0}\Omega_{0}=\Omega_{0}(I-\sum_{i}Z_{i}\otimes
dc_{i})=\Omega_{0}. \label{2.10f}%
\end{equation}

\begin{lemma}
\label{l4} If $\Omega$ is a presymplectic two-form compatible with a Poisson
bivector $\Pi_{0}$, then the bracket
\[
\{F,G\}^{\Omega}:=\Omega(X_{F}^{0},X_{G}^{0}),\ \ \ \ \ X_{F}^{0}=\Pi_{0}dF
\]
is a Poisson bracket.
\end{lemma}

\begin{proof}%
\begin{align*}
\{F,G\}^{\Omega}  &  =<\Omega X_{F}^{0},X_{G}^{0}>=<\Omega\Pi_{0}dF,\Pi
_{0}dG>=-<dG,\Pi_{0}\Omega\Pi_{0}dF>\\
&  =<dF,\Pi_{0}\Omega\Pi_{0}dG>=<dF,\Pi_{D}dG>\\
&  =\{F,G\}_{\Pi_{D}}%
\end{align*}
and $\Pi_{D}$ is Poisson.
\end{proof}

Obviously, when $\Omega=\Omega_{0},$ i.e. the compatible pair is simply a dual
pair, then we deal with a special case described by Lemma \ref{l3}. Moreover,
if $(\Pi,\Omega_{0})$ is a compatible P-p pair and $\ker(\Omega_{0}%
)=Sp\{Z_{i}\}_{i=1..r},$ then
\begin{equation}
\Omega_{0}(L_{Z_{i}}\Pi)\Omega_{0}=0,\ \ \ i=1,...,r, \label{d}%
\end{equation}
which follows from (\ref{1.2a}).

\begin{theorem}
\label{t2}Let $(\Pi_{0},\Omega_{0})$ be a dual P-p pair, such that $\ker
\Omega_{0}=Sp\{Z_{i}\}$ and $\ker\Pi_{0}=Sp\{dc_{i}\}.$ Moreover, let $\Pi$ be
a Poisson bivector compatible with $\Omega_{0},$ then:

\begin{description}
\item[(i)]
\begin{equation}
\Pi_{d}:=\Pi_{0}\Omega_{D}\Pi_{0}=\Pi_{0}\Omega_{0}\Pi\Omega_{0}\Pi
_{0}\nonumber
\end{equation}%
\begin{equation}
~\ \ \ \ \ \ \ \ \ \ \ \ \ \ \ \ \ \ =\Pi-\sum_{i}X_{i}\wedge Z_{i}+\frac
{1}{2}\sum_{i,j}c_{ij}Z_{i}\wedge Z_{j}, \label{2.8b}%
\end{equation}

\item[(ii)]
\begin{equation}
L_{Z_{i}}\Pi_{d}=0,\ \ \ i=1,...,r, \label{dd}%
\end{equation}

\item[(iii)]
\begin{equation}
L_{Z_{l}}\Pi=\sum_{i}[Z_{l},X_{i}]\wedge Z_{i}-\frac{1}{2}\sum_{i,j}%
Z_{l}(c_{ij})Z_{i}\wedge Z_{j}, \label{2.8a}%
\end{equation}
where $X_{i}=\Pi dc_{i},~c_{ij}=\Pi(dc_{i},dc_{j})=<dc_{i},\Pi dc_{j}>,$

\item[(iv)] $\Pi_{d}$ is Poisson.
\end{description}
\end{theorem}

\begin{proof}
From the definition of $\Pi_{d}$ we have%
\begin{align*}
\Pi_{d} &  =\Pi_{0}\Omega_{0}\Pi\Omega_{0}\Pi_{0}=(I-\sum_{i}Z_{i}\otimes
dc_{i})\Pi(I-\sum_{j}dc_{j}\otimes Z_{j})\\
&  =\Pi-\sum_{i}X_{i}\wedge Z_{i}+\frac{1}{2}\sum_{i,j}c_{ij}Z_{i}\wedge
Z_{j}.
\end{align*}
Then, from Lemma 3 and relation (\ref{d}), it follows that $L_{Z_{i}}\Pi
_{d}=0.$ Next, from (i) and (ii) immediately follows (iii). Finally we prove
the property (iv). If $X,Y$ are some vector fields, then their Schouten
bracket $[X,Y]_{S}=[X,Y]=L_{X}Y$ is a usual Lie bracket (commutator).
Moreover, for arbitrary bivector $P$ and function $F,$ the Schouten bracket
fulfills the relations
\begin{equation}
\lbrack X\wedge Y,P]_{S}=Y\wedge\lbrack X,P]_{S}-X\wedge\lbrack Y,P]_{S}%
,\ \ \ \ [X,P]_{S}=L_{X}P\label{0.36a}%
\end{equation}
and
\begin{equation}
L_{FX}P=FL_{X}P-(PdF)\wedge X.\label{2.5d}%
\end{equation}
Now, using (\ref{0.36a}) and (\ref{2.5d}), after straightforward but lengthy
calculations, one finds
\begin{align*}
\lbrack\Pi_{d},\Pi_{d}]_{S} &  =[\Pi,\Pi]_{S}-2[\Pi,\sum_{i}X_{i}\wedge
Z_{i}]_{S}+[\Pi,\sum_{i,j}c_{ij}Z_{i}\wedge Z_{j}]_{S}\\
&  +[\sum_{i}X_{i}\wedge Z_{i},\sum_{j}X_{j}\wedge Z_{j}]_{S}-[\sum_{k}%
X_{k}\wedge Z_{k},\sum_{i,j}c_{ij}Z_{i}\wedge Z_{j}]_{S}\\
&  +\frac{1}{4}[\sum_{i,j}c_{ij}Z_{i}\wedge Z_{j},\sum_{k,l}c_{kl}Z_{k}\wedge
Z_{l}]_{S}\\
&  =\sum_{i,j,k}X_{k}(c_{ij})Z_{i}\wedge Z_{k}\wedge Z_{j}=0,
\end{align*}
as
\[
\sum_{i,j,k}X_{k}(c_{ij})Z_{i}\wedge Z_{j}\wedge Z_{k}=\frac{1}{3}\sum
_{i,j,k}[X_{k}(c_{ij})+X_{k}(c_{ij})+X_{k}(c_{ij})]Z_{i}\wedge Z_{k}\wedge
Z_{j}=0
\]
which follows from Jacobi identity.
\end{proof}

As the concept of compatibility will be important in the reduction scheme for
bi-Hamiltonian chains, the following Theorem will be useful in the further considerations.

\begin{theorem}
\label{t1}Let $(\Pi_{0},\Omega_{0})$ be a dual P-p pair such that $\ker
\Omega_{0}=Sp\{Z_{i}\}$ and $\Pi$ be a Poisson tensor compatible with $\Pi
_{0}$. Then, $\Pi$ is compatible with $\Omega_{0}$ if
\begin{equation}
\Omega_{0}(L_{Z_{i}}\Pi)\Omega_{0}=0,\ \ \ i=1,...,k.\label{2.1}%
\end{equation}

\end{theorem}

\begin{proof}
First we gather all necessary formulas important for the calculation. For any
Poisson operator $\Pi$
\begin{equation}
L_{\Pi\gamma}\Pi=-\Pi(d\gamma)\Pi,\ \ \ \forall\gamma\in T^{\ast}M,
\label{2.2}%
\end{equation}
for any presymplectic form $\Omega$
\begin{equation}
L_{X}\Omega=d(\Omega X),\ \ \ \forall X\in TM \label{2.3}%
\end{equation}
and for an arbitrary second order mixed rank tensor $\Phi$%
\begin{equation}
\lbrack\Phi X_{1},X_{2}]=\Phi\lbrack X_{1},X_{2}]+(L_{X_{2}}\Phi)X_{1}.
\label{2.4}%
\end{equation}
For arbitrary vectors $X_{1},X_{2},X$ one-forms $\alpha_{1},\alpha_{2},$
two-form $\Omega,$ and function $F$ the following relations hold
\[
(X_{1}\otimes X_{2})(\alpha_{1}\otimes\alpha_{2})=\alpha_{1}(X_{2}%
)X_{1}\otimes\alpha_{2},\ \ \ \alpha_{1}(X_{2})=<\alpha_{1},X_{2}>,
\]%
\[
\Pi(\alpha_{1}\otimes\alpha_{2})=\Pi(\alpha_{1})\otimes\alpha_{2}%
,\ \ \ \ \Omega(X_{1}\otimes X_{2})=\Omega(X_{1})\otimes X_{2},
\]%
\[
(\alpha_{1}\otimes\alpha_{2})\Pi=-\alpha_{1}\otimes(\Pi\alpha_{2}%
),\ \ \ (X_{1}\otimes X_{2})\Omega=-X_{1}\otimes(\Omega X_{2}),
\]%
\begin{equation}
L_{FX}\Omega=FL_{X}\Omega+dF\wedge\Omega X. \label{2.5e}%
\end{equation}
As $\Pi_{0}$ and $\Pi$ are compatible so $\Pi+\lambda\Pi_{0}$ is Poisson,
hence for $\forall\tau\in TM$ and $\gamma=\Omega_{0}\tau$ from (\ref{2.2}) we
have
\begin{align*}
0  &  =L_{(\Pi+\lambda\Pi_{0})\gamma}(\Pi+\lambda\Pi_{0})+(\Pi+\lambda\Pi
_{0})d\gamma(\Pi+\lambda\Pi_{0})\\
&  =\lambda(L_{\Pi\gamma}\Pi_{0}+L_{\Pi_{0}\gamma}\Pi+\Pi(d\gamma)\Pi_{0}%
+\Pi_{0}(d\gamma)\Pi).
\end{align*}
Applying (\ref{partition}), (\ref{2.2}) and (\ref{2.5d}) we find
\[
L_{\Pi\gamma}\Pi_{0}=-\Pi_{0}(L_{\Pi\Omega_{0}\tau}\Omega_{0})\Pi_{0}-\sum
_{i}(\Pi_{0}da_{\gamma}^{i})\wedge Z_{i},
\]
where $a_{\gamma}^{i}=<dc_{i},\Pi\gamma>,L_{Z_{i}}\Omega_{0}=0$ and%
\[
L_{\Pi_{0}\gamma}\Pi=L_{\tau}\Pi-\sum_{i}L_{\tau(c_{i})Z_{i}}\Pi,
\]
hence
\[
0=-\Pi_{0}(L_{\Pi\Omega_{0}\tau}\Omega_{0})\Pi_{0}+\sum_{i}L_{a_{\gamma}%
^{i}Z_{i}}\Pi_{0}+L_{\tau}\Pi-\sum_{i}L_{\tau(c_{i})Z_{i}}\Pi+\Pi(L_{\tau
}\Omega_{0})\Pi_{0}+\Pi_{0}(L_{\tau}\Omega_{0})\Pi.
\]
Multiplying from left and right by $\Omega_{0}$ and using (\ref{partition}),
after strenuous but straightforward calculations with the application of
formulas (\ref{2.2})-(\ref{2.5e}) we arrive at the relation
\[
0=-d(\Omega_{0}\Pi\Omega_{0}\tau)+L_{\tau}(\Omega_{0}\Pi\Omega_{0})-\sum
_{i}[\Omega_{0}(L_{Z_{i}}\Pi)\Omega_{0}]\tau\wedge dc_{i}-\sum_{i}\tau
(c_{i})\Omega_{0}(L_{Z_{i}}\Pi)\Omega_{0}.
\]
Hence, $\Omega_{0}\Pi\Omega_{0}$ is closed if
\[
\sum_{i}[\Omega_{0}(L_{Z_{i}}\Pi)\Omega_{0}]\tau\wedge dc_{i}+\sum_{i}%
\tau(c_{i})\Omega_{0}(L_{Z_{i}}\Pi)\Omega_{0}=0.
\]
As the last equality holds for an arbitrary vector field $\tau,$ hence
\[
\Omega_{0}(L_{Z_{i}}\Pi)\Omega_{0}=0,\ \ \ i=1,...,r.
\]

\end{proof}

\begin{definition}
\label{d4}Let $(\Pi_{0},\Omega_{0})$ be a dual P-p pair and $\Pi$ be a Poisson
bivector. We say that $\Pi$ is compatible with the pair $(\Pi_{0},\Omega_{0})$
if $\Pi$ is compatible with $\Pi_{0}$ and $\Omega_{0}.$
\end{definition}

Up to now, we have induced a Poisson bracket on $C^{\infty}(\mathcal{M})$ in
various ways using not only Poisson bivectors but also dual pairs and
compatible pairs. So, the question is what is the most general way of
introducing a Poisson algebra on $C^{\infty}(\mathcal{M})$.

\begin{definition}
\label{d5}Assume that $\Pi$ is some bivector and $\Omega$ is a two-form. A
pair $(\Pi,\Omega)$ is called a Poisson pair if $\Pi_{D}=\Pi\Omega\Pi$ is
Poisson. Two Poisson pairs $(\Pi_{1},\Omega_{1})$ and $(\Pi_{2},\Omega_{2})$
will be called equivalent if $\Pi_{1}\Omega_{1}\Pi_{1}=\Pi_{2}\Omega_{2}%
\Pi_{2}.$
\end{definition}

Each compatible pair is simultaneously a Poisson pair. For a given Poisson
pair $(\Pi,\Omega)$ the bracket
\[
\{F,G\}_{\Pi}^{\Omega}:=\Omega(\Pi dF,\Pi dG)=<\Omega\Pi dF,\Pi dG>=<dF,\Pi
\Omega\Pi dG>
\]%
\[
=(\Pi\Omega\Pi)(dF,dG)=\{F,G\}_{\Pi_{D}}%
\ \ \ \ \ \ \ \ \ \ \ \ \ \ \ \ \ \ \ \ \ \
\]
is a Poisson bracket. Hence, the property of closeness of $\Omega$ is too
strong for the definition of a Poisson algebra.

\begin{definition}
\label{d6}Let $\Pi$ be a bivector with a kernel spanned by exact one-forms. A
two-form $\Omega$ is called weakly presymplectic with respect to $\Pi$ if it
is closed on $\operatorname{Im}\Pi=T\mathcal{N}$, where $\mathcal{N}$ is the
foliation given by functions whose differentials span the kernel of $\Pi$.
\end{definition}

Obviously, if $(\Pi,\Omega)$ is a Poisson pair then $\Omega$ is weakly
presymplectic with respect to $\Pi$. As we will see later, weakly
presymplectic forms play an important role in bi-Hamiltonian chains and in the
reduction procedure.

\section{Presymplectic representation of Gel'fand-\newline-Zakharevich chains}

Let us consider a bi-Poisson manifold $(M,\Pi_{0},\Pi_{1})$ of $\dim M=m=2n+r$
where $\Pi_{0},\Pi_{1}$ is a pair of compatible Poisson tensors of rank $2n.$
Moreover we assume that the Poisson pencil $\Pi_{\lambda}$ admits $r,$
polynomial with respect to the pencil parameter $\lambda,$ Casimir functions
of the form
\begin{equation}
H^{(j)}(\lambda)=\sum_{i=0}^{n_{j}}H_{i}^{(j)}\lambda^{n_{j}-i}%
,\ \ \ \ \ \ \ \ \ \ \ j=1,...,r, \label{0.4}%
\end{equation}
such that $n_{1}+...+n_{r}=n$ and $H_{i}^{(j)}$ are functionally independent.
The collection of $n$ bi-Hamiltonian vector fields
\begin{equation}
X_{i}^{(j)}=\Pi_{1}dH_{i-1}^{(j)}=\Pi_{0}dH_{i}^{(j)},\ \ \ \ i=1,...,n_{j}%
,\ \ \ j=1,...,r, \label{0.5}%
\end{equation}
constructed from Casimirs of the pencil
\[
\Pi_{\lambda}dH^{(j)}(\lambda)=0,
\]
is called the Gel'fand-Zakharevich system of the bi-Poisson manifold
$\mathcal{M}$ \cite{GZ},\cite{GZ1}. Notice that each chain starts from a
Casimir of $\Pi_{0}$ and terminates with a Casimir of $\Pi_{1}.$ Moreover all
$H_{i}^{(j)}$ pairwise commute with respect to both Poisson structures
\[
X_{i}^{(j)}(H_{l}^{(k)})=\langle dH_{l}^{(k)},\Pi_{0}dH_{i}^{(j)}%
\rangle=\langle dH_{l}^{(k)},\Pi_{1}dH_{i-1}^{(j)}\rangle=0.
\]%
\[
\Updownarrow
\]%
\[
\Pi_{\lambda}(dH_{i}^{(j)},dH_{l}^{(k)})=0.
\]

\subsection{Bi-presymplectic representation of one-Casimir chains}

As in this subsection we restrict our considerations to the simplest case of
$r=1,$ i.e. to the one-Casimir case, we will use the following notation for a
single bi-Hamiltonian chain
\begin{equation}
X_{i}=\Pi_{0}dH_{i}=\Pi_{1}dH_{i-1},\ \ \ \ \ \ i=0,...,n+1. \label{3.7}%
\end{equation}
The chain starts with a Casimir $H_{0}$ of $\Pi_{0}$ and terminates with a
Casimir $H_{n}$ of $\Pi_{1}.$

Let $\Omega_{0}$ be a dual to $\Pi_{0}$ presymplectic form. The kernels of
$\Omega_{0}$ and $\Pi_{0}$ are one dimensional: $\ker\Omega_{0}=Z$,\ $\ker
\Pi_{0}=dH_{0}$ and
\[
L_{Z}\Omega_{0}=0,\ \ \ L_{Z}\Pi_{0}=0.
\]
We assume that $\Omega_{0}(L_{Z}\Pi_{1})\Omega_{0}=0,$ i.e. that \ $\Pi_{1}$
is compatible with the P-p pair $(\Pi_{0},\Omega_{0}),$ so%
\[
L_{Z}\Pi_{1}=[Z,X_{1}]\wedge Z,\ \ \ \ X_{1}=\Pi_{1}dH_{0}%
\]
and
\[
\Omega_{1D}:=\Omega_{0}\Pi_{1}\Omega_{0}%
\]
is also presymplectic with $\ker\Omega_{0}\subseteq\ker\Omega_{1D}.$

Next, we construct the following two-form
\[
\Omega_{1}=\Omega_{1D}+\Omega_{0}X_{1}\wedge dH_{0}=\Omega_{1D}+dH_{1}\wedge
dH_{0}.
\]
It is obviously a presymplectic form. Moreover, $(\Pi_{0},\Omega_{1})$ is a
Poisson pair. Indeed,
\begin{equation}
\Pi_{0}\Omega_{1}\Pi_{0}=\Pi_{0}\Omega_{1D}\Pi_{0}+\Pi_{0}(dH_{1}\wedge
dH_{0})\Pi_{0}=\Pi_{0}\Omega_{1D}\Pi_{0}=\Pi_{1d}=\Pi_{1}-X_{1}\wedge Z
\label{3.10a}%
\end{equation}
which is Poisson according to Theorem \ref{t2}.

\begin{lemma}
\label{l5}Vector fields $Y=X_{n}+Z(H_{n})Z$ belong to $\ker\Omega_{1}.$
\end{lemma}

\begin{proof}%
\begin{align*}
\Omega_{1}Y &  =(\Omega_{1D}-dH_{0}\wedge dH_{1})(X_{n}+Z(H_{n})Z)\\
&  =(\Omega_{0}\Pi_{1}\Omega_{0})X_{n}-Z(H_{n})Z(H_{1})dH_{0}+Z(H_{n})dH_{1}.
\end{align*}
On the other hand, from (\ref{2.2a}) and the fact that $H_{0}$ is the only
Casimir function of $\Omega_{0}$
\begin{align*}
(\Omega_{0}\Pi_{1}\Omega_{0})X_{n} &  =\Omega_{0}\Pi_{1}(dH_{n}-Z(H_{n}%
)dH_{0})=-Z(H_{n})\Omega_{0}X_{1}\\
&  =-Z(H_{n})(dH_{1}-Z(H_{n})dH_{0})\\
&  =-Z(H_{n})dH_{1}+Z(H_{n})Z(H_{1})dH_{0}.
\end{align*}

\end{proof}

Now we are prepared to formulate the following theorem:

\begin{theorem}
\label{t3}Bi-presymplectic representation of the bi-Poisson chain (\ref{3.7})
takes the form
\begin{equation}
\beta_{i}=\Omega_{0}Y_{i}=\Omega_{1}Y_{i-1},\ \ \ i=0,...,n+1, \label{3.11}%
\end{equation}
where
\[
Y_{i}=X_{i}+Z(H_{i})Z,\ \ \ \ \beta_{i}=dH_{i}-Z(H_{i})dH_{0}.
\]
The chain starts with a kernel vector field $Y_{0}=Z$ of $\Omega_{0}$ and
terminates with a kernel vector field $Y_{n}\equiv Y=X_{n}+Z(H_{n})Z$ of
$\Omega_{1}.$
\end{theorem}

\begin{proof}%
\begin{align*}
\Omega_{0}Y_{i}  &  =\Omega_{0}X_{i},\\
\Omega_{1}Y_{i-1}  &  =(\Omega_{0}\Pi_{1}\Omega_{0}-dH_{0}\wedge
dH_{1})(X_{i-1}+Z(H_{i-1})Z)\\
&  =(\Omega_{0}\Pi_{1}\Omega_{0})X_{i-1}-Z(H_{i-1})Z(H_{1})dH_{0}%
+Z(H_{i-1})dH_{1},\\
(\Omega_{0}\Pi_{1}\Omega_{0})X_{i-1}  &  =\Omega_{0}\Pi_{1}(dH_{i-1}%
-Z(H_{i-1})dH_{0})=\Omega_{0}(X_{i}-Z(H_{i-1})X_{1}\\
&  =\Omega_{0}X_{i}-Z(H_{i-1})\Omega_{0}X_{1}\\
&  =\Omega_{0}X_{i}-Z(H_{i-1})dH_{1}+Z(H_{i-1})Z(H_{1})dH_{0}.
\end{align*}

\end{proof}

Observe that neither $X_{i}$ nor $Y_{i}$ vector fields are inverse Hamiltonian
with respect to $\Omega_{0}$ and $\Omega_{1}$. Besides $[Y_{i},Y_{j}]\neq0.$
Introducing a presymplectic pencil
\[
\Omega_{\lambda}=\Omega_{1}-\lambda\Omega_{0}%
\]
with a kernel vector field
\[
Y=\sum_{i=0}^{n}Y_{i}\lambda^{n-i},
\]
the bi-presymplectic chain (\ref{3.11}) takes the form $\Omega_{\lambda}Y=0.$
On the other hand, the pairs $(\Pi_{0},\Omega_{0})$ and $(\Pi_{0},\Omega_{1})$
are Poisson pairs, hence $\Omega_{0}$ and $\Omega_{1}$ define Poisson
brackets. The first one is equal to that given by $\Pi_{0}$ (\ref{2.10f})
while the second one is equal to that given by $\Pi_{1d}$ (\ref{3.10a}).
Moreover,
\[
\Omega_{0}(X_{i},X_{j})=\{H_{i},H_{j}\}_{\Pi_{0}}=0,\ \ \ \ \ \Omega_{1}%
(X_{i},X_{j})=\{H_{i},H_{j}\}_{\Pi_{1d}}=0.
\]
The first bracket is obvious, the second one follows from the relation
\begin{align*}
\Omega_{1}X_{i}  &  =(\Omega_{1D}+dH_{1}\wedge dH_{0})X_{i}=\Omega_{0}\Pi
_{1}\Omega_{0}X_{i}=\Omega_{0}\Pi_{1}(dH_{i}-Z(H_{i})dH_{0}\\
&  =\Omega_{0}X_{i+1}-Z(H_{i})\Omega_{0}X_{1}%
\end{align*}
and the first bracket. Additionally, Poisson tensors $\Pi_{0}$ and $\Pi_{1d}$
are compatible as
\[
\lbrack\Pi_{1d},\Pi_{0}]_{S}=[\Pi_{1}-X_{1}\wedge Z,\Pi_{0}]_{S}=X_{1}%
\wedge\lbrack Z,\Pi_{0}]_{S}-Z\wedge\lbrack X_{1},\Pi_{0}]_{S}=0.
\]
As a consequence $(\Pi_{0},\Omega_{\lambda})$ is a Poisson pair and
\[
\Omega_{\lambda}(X_{i},X_{j})=0.
\]

\subsection{Weakly bi-presymplectic representation of multi-Casimir chains}

In this subsection we will show that bi-presymplectic representation is purely
one-Casimir phenomenon. Consider the $r$-Casimir Gel'fand-Zakharevich chain
(\ref{0.4}),(\ref{0.5}). Let $\Omega_{0}$ be a dual to $\Pi_{0}$ presymplectic
form. The kernels of $\Omega_{0}$ and $\Pi_{0}$ are $r$ dimensional:
$\ker\Omega_{0}=Sp\{Z_{i}\}_{i=1,...,r}$,\ $\ker\Pi_{0}=Sp\{dH_{0}%
^{(i)}\}_{i=1,...,r}$ and
\begin{equation}
L_{Z_{i}}\Omega_{0}=0,\ \ \ L_{Z_{i}}\Pi_{0}=0,\ \ \ \ i=1,...,r. \label{3.16}%
\end{equation}
We assume that $\Omega_{0}(L_{Z_{i}}\Pi_{1})\Omega_{0}=0,$ i.e. that
\ $\Pi_{1}$ is compatible with the P-p pair $(\Pi_{0},\Omega_{0}),$ so from
involutivity of $H_{k}^{(i)}$ the relation (\ref{2.8a}) takes the form%
\[
L_{Z_{i}}\Pi_{1}=\sum_{k}[Z_{i},X_{1}^{(k)}]\wedge Z_{k},\ \ \ \ X_{1}%
^{(k)}=\Pi_{1}dH_{0}^{(k)}%
\]
and
\[
\Omega_{1D}:=\Omega_{0}\Pi_{1}\Omega_{0}%
\]
is also presymplectic with $\ker\Omega_{0}\subseteq\ker\Omega_{1D}.$

Next, we construct the following two-forms
\[
\overline{\Omega}_{1}=\Omega_{1D}+\sum_{j=1}^{r}\Omega_{0}X_{1}^{(j)}\wedge
dH_{0}^{(j)},\ \ \ \ \ \Omega_{1}=\Omega_{1D}+\sum_{j=1}^{r}dH_{1}^{(j)}\wedge
dH_{0}^{(j)},
\]
related with each other as follows
\[
\Omega_{1}=\overline{\Omega}_{1}+\frac{1}{2}\sum_{k,l}A_{kl}dH_{0}^{(k)}\wedge
dH_{0}^{(l)},\ \ \ \ \ A_{kl}=Z_{k}(H_{1}^{(l)})-Z_{l}(H_{1}^{(k)}).
\]
Obviously $\Omega_{1}$ is\ presymplectic and together with $\Pi_{0}$ forms a
Poisson pair as
\[
\Pi_{0}\Omega_{1}\Pi_{0}=\Pi_{0}\Omega_{1D}\Pi_{0}=\Pi_{0}\Omega_{0}\Pi
_{1}\Omega_{0}\Pi_{0}=\Pi_{1d}=\Pi_{1}-\sum_{i}X_{1}^{(i)}\wedge Z_{i}%
\]
is Poisson. It is also clear that $\overline{\Omega}_{1}$\ is not closed as
\[
d\overline{\Omega}_{1}=-\frac{1}{2}\sum_{k,l}dA_{kl}\wedge dH_{0}^{(k)}\wedge
dH_{0}^{(l)},
\]
but is weakly presymplectic with respect to $\Pi_{0}$
\[
d\overline{\Omega}_{1}(\Pi_{0}\alpha_{1},\Pi_{0}\alpha_{2},\Pi_{0}\alpha
_{3})=0,\ \ \ \ \ \ \ \forall\alpha_{1},\alpha_{2},\alpha_{3}\in T^{\ast
}\mathcal{M}.
\]
Moreover, $(\Pi_{0},\overline{\Omega}_{1})$ is a Poisson pair equivalent to
$(\Pi_{0},\Omega_{1})$ one as $\Pi_{0}\overline{\Omega}_{1}\Pi_{0}=\Pi
_{0}\Omega_{1}\Pi_{0}=\Pi_{1d}$.

\begin{theorem}
\label{t5}Multi-Casimir bi-Poisson chains (\ref{0.5}) have weakly
bi-presymplectic representation
\begin{equation}
\beta_{i}^{(j)}=\Omega_{0}Y_{i}^{(j)}=\overline{\Omega}_{1}Y_{i-1}%
^{(j)},\ \ \ j=1,...,r,\ \ i=0,...,n_{j}+1, \label{3.23}%
\end{equation}
where
\[
Y_{i}^{(j)}=X_{i}^{(j)}+\sum_{k=1}^{r}Z_{k}(H_{i}^{(j)})Z_{k},\ \ \ \beta
_{i}^{(j)}=dH_{i}^{(j)}-\sum_{k=1}^{r}Z_{k}(H_{i}^{(j)})dH_{0}^{(k)}.
\]
The $j$-th chain starts with a kernel vector field $Y_{0}^{(j)}=Z_{j}$ of
$\Omega_{0}$ and terminates with a kernel vector field $Y_{n_{j}}%
^{(j)}=X_{n_{j}}^{(j)}+\sum_{k=1}^{m}Z_{k}(H_{n_{j}}^{(j)})Z_{k}$ of
$\overline{\Omega}_{1}.$
\end{theorem}

\begin{proof}
We have%
\[
\Omega_{0}Y_{i}^{(j)}=\Omega_{0}X_{i}^{(j)}.
\]
On the other hand
\begin{align*}
\overline{\Omega}_{1}Y_{i-1}^{(j)}  &  =(\Omega_{0}\Pi_{1}\Omega_{0}+\sum
_{l}\Omega_{0}X_{1}^{(l)}\wedge dH_{0}^{(l)})(X_{i-1}^{(j)}+\sum_{k}%
Z_{k}(H_{i-1}^{(j)})Z_{k})\\
&  =\Omega_{0}\Pi_{1}\Omega_{0}X_{i-1}^{(j)}+(\sum_{l}\Omega_{0}X_{1}%
^{(l)}\wedge dH_{0}^{(l)})X_{i-1}^{(j)}\\
&  +\sum_{l,k}Z_{k}(H_{i-1}^{(j)})(\Omega_{0}X_{1}^{(l)}\wedge dH_{0}%
^{(l)})Z_{k}.
\end{align*}
Using decomposition (\ref{2.2a}) and bi-Hamiltonian chains (\ref{0.5}) one
finds%
\[
\Omega_{0}\Pi_{1}\Omega_{0}X_{i-1}^{(j)}=\Omega_{0}X_{i}^{(j)}-\sum_{k}%
Z_{k}(H_{i-1}^{(j)})dH_{1}^{(k)}+\sum_{l,k}Z_{k}(H_{i-1}^{(j)})Z_{l}%
(H_{1}^{(k)})dH_{0}^{(l)},
\]%
\[
\sum_{l,k}Z_{k}(H_{i-1}^{(j)})(\Omega_{0}X_{1}^{(l)}\wedge dH_{0}^{(l)}%
)Z_{k}=\sum_{k}Z_{k}(H_{i-1}^{(j)})dH_{1}^{(k)}-\sum_{j,k}Z_{k}(H_{i-1}%
^{(j)})Z_{l}(H_{1}^{(k)})dH_{0}^{(l)},
\]
\[
(\sum_{l}\Omega_{0}X_{1}^{(l)}\wedge dH_{0}^{(l)})X_{i-1}^{(j)}=-\sum
_{l}\Omega_{0}(X_{1}^{(l)},X_{i-1}^{(j)})dH_{0}^{(l)}=0.
\]
The last equality follows from the fact that $\Omega_{0}(X_{1}^{(l)}%
,X_{i-1}^{(j)})=\Pi_{0}(dH_{1}^{(l)},dH_{i-1}^{(j)})=0.$ Hence
\[
\overline{\Omega}_{1}Y_{i-1}^{(j)}=\Omega_{0}X_{i}^{(j)}.
\]

\end{proof}

Introducing a weakly presymplectic pencil
\[
\overline{\Omega}_{\lambda}=\overline{\Omega}_{1}-\lambda\Omega_{0}%
\]
with respect to $\Pi_{0},$ with a kernel vector fields
\[
Y^{(j)}=\sum_{i=0}^{n_{j}}Y_{i}^{(j)}\lambda^{n_{j}-i},\ \ \ \ j=1,...,r,
\]
the weakly bi-presymplectic chains (\ref{3.23}) take the form $\overline
{\Omega}_{\lambda}Y^{(j)}=0.$ On the other hand, as we mentioned before, the
pairs $(\Pi_{0},\Omega_{0})$ and $(\Pi_{0},\overline{\Omega}_{1})$ are Poisson
pairs, hence $\Omega_{0}$ and $\overline{\Omega}_{1}$ define Poisson brackets.
The first one is equal to that given by $\Pi_{0}$ while the second one is
equal to that given by $\Pi_{1d}$. Moreover,
\[
\Omega_{0}(X_{i}^{(k)},X_{j}^{(l)})=\{H_{i}^{(k)},H_{j}^{(l)}\}_{\Pi_{0}%
}=0,\ \ \Omega_{1}(X_{i}^{(k)},X_{j}^{(l)})=\{H_{i}^{(k)},H_{j}^{(l)}%
\}_{\Pi_{1d}}=0.
\]
The first bracket is obvious, the second one follows from the relation
\begin{align*}
\Omega_{1}X_{i}^{(k)}  &  =(\Omega_{1D}+\sum_{r}dH_{1}^{(r)}\wedge
dH_{0}^{(r)})X_{i}^{(k)}=\Omega_{0}\Pi_{1}\Omega_{0}X_{i}^{(k)}\\
&  =\Omega_{0}\Pi_{1}(dH_{i}^{(k)}-\sum_{r}Z_{r}(H_{i}^{(k)})dH_{0}^{(r)})\\
&  =\Omega_{0}X_{i+1}^{(k)}-\sum_{r}Z_{r}(H_{i}^{(k)})\Omega_{0}X_{1}^{(r)}%
\end{align*}
and the first bracket. Additionally, Poisson tensors $\Pi_{0}$ and $\Pi_{1d}$
are compatible as
\begin{align*}
\lbrack\Pi_{1d},\Pi_{0}]_{S}  &  =[\Pi_{1}-\sum_{i}X_{1}^{(i)}\wedge Z_{i}%
,\Pi_{0}]_{S}\\
&  =\sum_{i}(X_{1}^{(i)}\wedge\lbrack Z_{i},\Pi_{0}]_{S}-Z_{i}\wedge\lbrack
X_{1}^{(i)},\Pi_{0}]_{S})\\
&  =\sum_{i}(X_{1}^{(i)}\wedge L_{Z_{i}}\Pi_{0}-Z_{i}\wedge L_{X_{1}^{(i)}}%
\Pi_{0})\\
&  =0.
\end{align*}
As a consequence, $(\Pi_{0},\overline{\Omega}_{\lambda})$ is a Poisson pair
and
\[
\overline{\Omega}_{\lambda}(X_{i},X_{j})=0.
\]

Now, let us consider the presymplectic pencil
\[
\Omega_{\lambda}=\Omega_{1}-\lambda\Omega_{0}.
\]
As $(\Pi_{0},\Omega_{1})$ is a Poisson pair equivalent to the Poisson pair
$(\Pi_{0},\overline{\Omega}_{1}),$ then
\[
\Omega_{\lambda}(X_{i},X_{j})=0.
\]
Moreover, chains (\ref{3.23}) take the form
\[
\beta_{i}^{(j)}=\Omega_{0}Y_{i}^{(j)}=\Omega_{1}Y_{i-1}^{(j)}-\sum
_{k}B_{i-1,k}^{(j)}dH_{0}^{(k)},\ \ \ \ B_{i,k}^{(j)}=\sum_{l}A_{kl}%
Z_{l}(H_{i}^{(j)}),
\]
where $j=1,...,r,\ \ i=0,...,n_{j}+1.$

\section{Reduction procedure for Gel'fand-Zakharevich chains}

Let us consider a $(2n+r)$-dimensional manifold $\mathcal{M}$ and
$2n$-dimensional submanifold $\mathcal{N}$ of $\mathcal{M}$. Then, let us fix
an integrable distribution $\mathcal{Z}$ of constant dimension $r$ that is
transversal to $\mathcal{N}$. As mentioned in Section 2, such a case is
realized by an appropriate dual P-p pair defined on $\mathcal{M}$. Indeed, let
$(\Pi_{0},\Omega_{0})$ be a dual P-p pair on $\mathcal{M}$ with $\ker
\Omega_{0}=\mathcal{Z}=Sp\{Z_{i}\}$ and $\ker\Pi_{0}=\mathcal{Z}^{\ast
}=Sp\{dc_{i}\},\ i=1,...,r$ where obviously $Z_{i}(c_{j})=\delta_{ij\text{ }}$
and $[Z_{i},Z_{j}]=0.$ Then, $\mathcal{N}$ is a fixed symplectic leave of
$\Pi$ and $\mathcal{Z}$ consists of vector fields from $\ker\Omega_{0}$
evaluated on $\mathcal{N}$. An appropriate decomposition of tangent and
cotangent bundle of $\mathcal{M}$ is given by (\ref{2.1b}).

\begin{definition}
A function $F$ :$\ \mathcal{M}\rightarrow\mathbb{R}$ is called invariant with
respect to distribution $\mathcal{Z}$ if
\[
L_{Z_{i}}F=Z_{i}(F)=0,\ \ \ \forall Z_{i}\in\mathcal{Z}.
\]
The set of such functions will be denoted by $\mathcal{A}.$
\end{definition}

\begin{definition}
The Poisson tensor $\Pi$ is called invariant with respect to the distribution
$\mathcal{Z}$ if functions that are invariant along $\mathcal{Z} $ form a
Poisson subalgebra with respect to $\Pi$, that is
\begin{equation}
L_{Z_{i}}\Pi(dF,dG)=0,\ \ \ Z_{i}(F)=Z_{i}(G)=0. \label{3.2}%
\end{equation}
We will denote this subalgebra by $\mathcal{A}(\Pi).$
\end{definition}

Notice that $\Pi_{0}$ is in obvious way $\mathcal{Z}$-invariant as $L_{Z_{i}%
}\Pi_{0}=0,$ hence $\mathcal{A}(\Pi_{0})$ is also a Poisson subalgebra.

\begin{lemma}
\label{l6}If Poisson bivector $\Pi$ is compatible with a presymplectic form
$\Omega_{0},$ then it is invariant with respect to the distribution
$\mathcal{Z}=\ker\Omega_{0}$.
\end{lemma}

\begin{proof}
Assume $Z_{i}(F)=Z_{i}(G)=0$ for all $i$. We have to show that condition
(\ref{3.2}) is fulfilled. But due to Theorem \ref{t2} it follows that
\begin{align*}
L_{Z_{l}}\Pi(dF,dG)  &  =(L_{Z_{l}}\Pi)(dF,dG)=<dF,(L_{Z_{l}}\Pi)dG>\\
&  =<dF,\left(  \sum_{i}[Z_{l},X_{i}]\wedge Z_{i}-\frac{1}{2}\sum_{i,j}%
Z_{l}(c_{ij})Z_{i}\wedge Z_{j}\right)  dG>\\
&  =\sum_{i}\left(  Z_{i}(G)[Z_{l},X_{i}](F)-Z_{i}(F)[Z_{l},X_{i}](G)\right)
\\
&  -\frac{1}{2}\sum_{i,j}Z_{l}(c_{ij})[Z_{j}(G)Z_{i}(F)-Z_{j}(F)Z_{i}(G)]\\
&  =0.
\end{align*}

\end{proof}

The invariance of Poisson tensors given in the form (\ref{2.8b}) was proved
for the first time by Vaisman \cite{Vaisman}.

As a consequence we conclude that an arbitrary Poisson bivector $\Pi,$
compatible with a dual P-p pair $(\Pi_{0},\Omega_{0}),$ is reducible onto
foliation given by Casimirs of $\Pi_{0}$ along the distribution given by
$\ker\Omega_{0}$. Here we propose a simple constructive method of deriving the
reduced operator.

\begin{lemma}
\label{l7}Let $\Pi$ be a Poisson bivector compatible with a dual P-p pair
$(\Pi_{0},\Omega_{0})$ and $\pi$ a reduction of $\Pi$ onto a symplectic leaf
$\mathcal{N}_{\nu}$ of $\Pi_{0}$ along the transversal distribution
$\mathcal{Z}=\ker\Omega_{0}.$ Then, $\pi$ can be constructed by a restriction
of
\[
\Pi_{d}=\Pi_{0}\Omega_{0}\Pi\Omega_{0}\Pi_{0}=\Pi-\sum_{i}X_{i}\wedge
Z_{i}+\frac{1}{2}\sum_{i,j}c_{ij}Z_{i}\wedge Z_{j}%
\]
to $\mathcal{N}_{\nu}$%
\begin{equation}
\pi=\Pi_{d}|_{\mathcal{N}_{\nu}}. \label{red1}%
\end{equation}

\end{lemma}

\begin{proof}
From the relation (\ref{2.8b}) and the fact that for $F,G\in\mathcal{A}$
\[
<dF,(-\sum_{i}X_{i}\wedge Z_{i}+\frac{1}{2}\sum_{i,j}c_{ij}Z_{i}\wedge
Z_{j})dG>=0,
\]
the Poisson operator $\Pi$ and its deformation $\Pi_{d}$ both act in the same
way on the set $\mathcal{A}$, so that both can be used to define the same
reduced operator $\pi$ on $\mathcal{N}_{\nu}.$ But as the image of $\Pi_{d}$
is tangent to $\mathcal{N}_{\nu},$ what follows from the fact that $\ker
\Pi_{0}\subset\ker\Pi_{d}$,  and $\Pi_{d}$ is Poisson, then the projection of
$\Pi_{d}$ onto $\mathcal{N}_{\nu}$ means simple its restriction to
$\mathcal{N}_{\nu}.$ Obviously, if $\ker\Pi_{d}=\ker\Pi_{0}$, then
(\ref{red1}) means the restriction of $\Pi_{d}$ to its symplectic leaf
$\mathcal{N}_{\nu}$.
\end{proof}

Now we pass to the reduction of bi-Hamiltonian chains in Poisson (\ref{0.5})
and presymplectic (\ref{3.23}) representations onto symplectic foliation of
$\Pi_{0}$. Let us denote the projections of $\Pi_{0},\Pi_{1}$ onto
$\mathcal{N}$ along $\mathcal{Z}$ by $\pi_{0},\pi_{1}$ and restrictions of
$(H_{1}^{(1)},...,$ $H_{n_{r}}^{(r)})|_{\mathcal{N}}$ to $\mathcal{N}$ by
$(h_{1}^{(1)},...,h_{n_{r}}^{(r)}).$

\begin{proposition}
The bi-Poisson chain (\ref{0.5}), when reduced to $\mathcal{N}$ takes the
form
\begin{equation}
\pi_{1}dh_{i}^{(j)}=\pi_{0}dh_{i+1}^{(j)}-\sum_{k=1}^{r}\alpha_{ki}^{(j)}%
\pi_{0}dh_{1}^{(k)},\ \ \ \ \ j=1,...,r,\ \ i=1,...,n_{j}, \label{4.1}%
\end{equation}
where $\ \alpha_{ki}^{(j)}=Z_{k}(H_{i}^{(j)})|_{\mathcal{N}}.$
\end{proposition}

\begin{proof}%
\begin{align*}
\pi_{1}dh_{i}^{(j)}  &  =\Pi_{1d}|_{\mathcal{N}}~dH_{i}^{(j)}|_{\mathcal{N}%
}=(\Pi_{1d}~dH_{i}^{(j)})|_{\mathcal{N}}\\
&  =(\Pi_{1}dH_{i}^{(j)})|_{\mathcal{N}}-\sum_{k=1}^{r}\left(  Z_{k}%
(H_{i}^{(j)})X_{1}^{(k)}\right)  |_{\mathcal{N}}\\
&  =(\Pi_{0}dH_{i+1}^{(j)})|_{\mathcal{N}}-\sum_{k=1}^{r}\left(  Z_{k}%
(H_{i}^{(j)})\Pi_{0}dH_{1}^{(k)}\right)  |_{\mathcal{N}}\\
&  =\Pi_{0}|_{\mathcal{N}}\ dH_{i+1}^{(j)}|_{\mathcal{N}}-\sum_{k=1}^{r}%
Z_{k}(H_{i}^{(j)})|_{\mathcal{N}}\ \Pi_{0}|_{\mathcal{N}}\ dH_{1}%
^{(k)}|_{\mathcal{N}}\\
&  =\pi_{0}dh_{i+1}^{(j)}-\sum_{k=1}^{r}Z_{k}(H_{i}^{(j)})|_{\mathcal{N}}%
\pi_{0}dh_{1}^{(k)}.
\end{align*}
The second and fifth equalities are valid as in coordinates
\begin{equation}
(x^{i},H_{0}^{(j)}),\ \ \ \ \ \ \ i=1,...,2n,\ \ j=1,...,r \label{4.2}%
\end{equation}
on $\mathcal{M}$, the last $r$ rows and columns of $\Pi_{0}$ and $\Pi_{1d}$
contain zeros only. Obviously we have
\[
\pi_{0}(dh_{i}^{(j)},dh_{k}^{(l)})=\pi_{1}(dh_{i}^{(j)},dh_{k}^{(l)})=0,
\]
which follows from the construction of $\pi_{0}$ and $\pi_{1}.$
\end{proof}

Before we pass to the reduction of presymplectic representation (\ref{3.23}),
observe that as $(\Pi_{0},\Omega_{0}),$ $(\Pi_{0},\Omega_{1})$ and $(\Pi
_{0},\overline{\Omega}_{1})$ are Poisson pairs, then their restrictions to
$\mathcal{N}$ are closed: $\Omega_{0}|_{\mathcal{N}}=\omega_{0}=\pi
^{-1},~\Omega_{1}|_{\mathcal{N}}=\overline{\Omega}_{1}|_{\mathcal{N}}%
=\omega_{1}.$ Moreover, $\pi_{0}dh_{i}^{(j)}:=K_{i}^{(j)}=X_{i}^{(j)}%
|_{\mathcal{N}},$ where $|_{\mathcal{N}}$ means as usually a restriction, as
\[
X_{i}^{(j)}|_{\mathcal{N}}=(\Pi_{0}dH_{i}^{(j)})|_{\mathcal{N}}=\Pi
_{0}|_{\mathcal{N}}\ dH_{i}^{(j)}|_{\mathcal{N}}=\pi_{0}dh_{i}^{(j)}.
\]

\begin{proposition}
When reduced to $\mathcal{N}$ , the weakly bi-presymplectic chain (\ref{0.5})
takes the form
\begin{equation}
\omega_{1}K_{i}^{(j)}=\omega_{0}K_{i+1}^{(j)}-\sum_{k}\alpha_{ki}^{(j)}%
\omega_{0}K_{1}^{(j)},\ \ \ \ \ j=1,...,r,\ \ i=1,...,n_{j}. \label{4.3}%
\end{equation}

\end{proposition}

\begin{proof}%
\begin{align*}
\omega_{1}K_{i}^{(j)}  &  =\overline{\Omega}_{1}|_{\mathcal{N}}\ X_{i}%
^{(j)}|_{\mathcal{N}}=(\overline{\Omega}_{1}X_{i}^{(j)})|_{\mathcal{N}%
}=(\overline{\Omega}_{1}(Y_{i}^{(j)}-\sum_{k}Z_{k}(H_{i}^{(j)})Z_{k}%
))|_{\mathcal{N}}\\
&  =(\overline{\Omega}_{1}Y_{i}^{(j)})|_{\mathcal{N}}-\sum_{k}(Z_{k}%
(H_{i}^{(j)})\beta_{1}^{(j)})|_{\mathcal{N}}\\
&  =(\Omega_{0}Y_{i+1}^{(j)})|_{\mathcal{N}}-\sum_{k}(Z_{k}(H_{i}^{(j)}%
)\Omega_{0}Y_{1}^{(k)})|_{\mathcal{N}}\\
&  =(\Omega_{0}X_{i+1}^{(j)})|_{\mathcal{N}}-\sum_{k}(Z_{k}(H_{i}^{(j)}%
)\Omega_{0}X_{1}^{(k)})|_{\mathcal{N}}\\
&  =\Omega_{0}|_{\mathcal{N}}~X_{i+1}^{(j)}|_{\mathcal{N}}-\sum_{k}Z_{k}%
(H_{i}^{(j)})|_{\mathcal{N}}~\Omega_{0}|_{\mathcal{N}}~X_{1}^{(k)}%
|_{\mathcal{N}}\\
&  =\omega_{0}K_{i+1}^{(j)}-\sum_{k}\alpha_{ki}^{(j)}\omega_{0}K_{1}^{(j)}.
\end{align*}
The second and seventh equality are valid as in the coordinates (\ref{4.2})
vector fields $X_{i}^{(j)}$ have the last $r$ components equal to zero.
\end{proof}

Notice that
\begin{align*}
\omega_{1}  &  =\overline{\Omega}_{1}|_{\mathcal{N}}=\Omega_{1D}%
|_{\mathcal{N}}=(\Omega_{0}\Pi_{1}\Omega_{0})|_{\mathcal{N}}=(\Omega_{0}%
\Pi_{1d}\Omega_{0})|_{\mathcal{N}}\\
&  =\Omega_{0}|_{\mathcal{N}}\ \Pi_{1d}|_{\mathcal{N}}\ \Omega_{0}%
|_{\mathcal{N}}=\omega_{0}\pi_{1}\omega_{0}.
\end{align*}
As $\omega_{1}$ is closed then $(\pi_{1},\omega_{0})$ is a compatible pair and
$N=\pi_{1}\omega_{0}$ is a recursion operator. Moreover $\pi_{1}=N\pi_{0}$
hence $\pi_{0}$ and $\pi_{1}$ are compatible. Now we immediately find that
reduced chains (\ref{4.1}) and (\ref{4.3}) are equivalent. As $K_{i}^{(j)}%
=\pi_{0}dh_{i}^{(j)},$ hence (\ref{4.3}) takes the form
\begin{equation}
N^{\ast}dh_{i}^{(j)}=dh_{i+1}^{(j)}-\sum_{k}\alpha_{ki}^{(j)}dh_{1}%
^{(j)},\ \ \ \ \ j=1,...,r,\ \ i=1,...,n_{j}, \label{4.4}%
\end{equation}
where $N^{\ast}=\omega_{0}\pi_{1}$ is a recursion operator for one-forms. On
the other hand, multiplying (\ref{4.1}) from left by $\omega_{0}$ we arrive at
(\ref{4.4}) again. Moreover,
\[
\omega_{0}(K_{i}^{(j)},K_{l}^{(r)})=\pi_{0}(dh_{i}^{(j)},dh_{k}^{(l)}%
)=0,\ \ \ \omega_{1}(K_{i}^{(j)},K_{l}^{(r)})=\pi_{1}(dh_{i}^{(j)}%
,dh_{k}^{(l)})=0.
\]
As a consequence, the distribution tangent to the foliation of $\mathcal{N}$
defined by $(h_{1}^{(1)},...,h_{n_{r}}^{(r)})$ is bi-Lagrangian and the
$n-$tuple $(h_{1}^{(1)},...,h_{n_{r}}^{(r)})$ of functionally independent
Hamiltonians is separable \cite{m3}. Separated coordinates are eigenvalues of
the recursion operator $N$ and canonically conjugated momenta that put the
recursion operator in the diagonal form.

We conclude this section with a statement, that the existence of weakly
bi-presymplectic representation of bi-Poisson chains is a sufficient condition
for the separability of related Hamiltonian systems.

\section{Example}

Let us illustrate our previous considerations with a simple nontrivial example
of the integrable case of the Henon-Heiles equations
\begin{align}
(q^{1})_{tt}  &  =-3(q^{1})^{2}-\frac{1}{2}(q^{2})^{2}+c\nonumber\\
(q^{2})_{tt}  &  =-q^{1}q^{2}. \label{5.0}%
\end{align}
The system (\ref{5.0}) can be put into a canonical Hamiltonian form with the
Hamiltonian function given by
\[
H_{1}=\frac{1}{2}p_{1}^{2}+\frac{1}{2}p_{2}^{2}+(q^{1})^{3}+\frac{1}{2}%
q^{1}(q^{2})^{2}-cq^{1},
\]
where $p_{1}=q_{t}^{1},p_{2}=q_{t}^{2}.$The second constant of motion is
\[
H_{2}=\frac{1}{2}q^{2}p_{1}p_{2}-\frac{1}{2}q^{1}p_{2}^{2}+\frac{1}{16}%
(q^{2})^{4}+\frac{1}{4}(q^{1})^{2}(q^{2})^{2}-\frac{1}{4}c(q^{2})^{2}.
\]
The bi-Hamiltonian chain on $\mathcal{M}=Sp(q^{1},q^{2},p_{1},p_{2},c)$ is of
the following form
\[%
\begin{array}
[c]{l}%
\Pi_{0}dH_{0}=0\\
\Pi_{0}dH_{1}=X_{1}=\Pi_{1}dH_{0}\\
\Pi_{0}dH_{2}=X_{2}=\Pi_{1}dH_{1}\\
\qquad\qquad\,\,\ \,\,\,0=\Pi_{1}dH_{2,}\,\,
\end{array}
\]
where $H_{0}=c$ and the compatible Poisson bivectors are%
\[
\Pi_{0}=\left(
\begin{array}
[c]{ccccc}%
0 & 0 & 1 & 0 & 0\\
0 & 0 & 0 & 1 & 0\\
-1 & 0 & 0 & 0 & 0\\
0 & -1 & 0 & 0 & 0\\
0 & 0 & 0 & 0 & 0
\end{array}
\right)
,\ \ \ \ \ \ \ \ \ \ \ \ \ \ \ \ \ \ \ \ \ \ \ \ \ \ \ \ \ \ \ \ \ \ \ \ \ \ \ \ \ \ \ \ \ \ \ \ \ \ \ \ \ \ \ \ \ \ \ \ \ \ \ \ \ \ \ \ \ \ \ \ \ \ \ \ \ \ \ \text{\ }%
\]%
\[
\Pi_{1}=\left(
\begin{array}
[c]{ccccc}%
0 & 0 & q^{1} & \frac{1}{2}q^{2} & p_{1}\\
0 & 0 & \frac{1}{2}q^{2} & 0 & p_{2}\\
-q^{1} & -\frac{1}{2}q^{2} & 0 & \frac{1}{2}p_{2} & -3(q^{1})^{2}-\frac{1}%
{2}(q^{2})^{2}+c\\
-\frac{1}{2}q^{2} & 0 & -\frac{1}{2}p_{2} & 0 & -q^{1}q^{2}\\
-p_{1} & -p_{2} & 3(q^{1})^{2}+\frac{1}{2}(q^{2})^{2}-c & q^{1}q^{2} & 0
\end{array}
\right)  .
\]
Now, dual to the canonical Poisson tensor $\Pi_{0}$ is a canonical
presymplectic form
\[
\Omega_{0}=\left(
\begin{array}
[c]{ccccc}%
0 & 0 & -1 & 0 & 0\\
0 & 0 & 0 & -1 & 0\\
1 & 0 & 0 & 0 & 0\\
0 & 1 & 0 & 0 & 0\\
0 & 0 & 0 & 0 & 0
\end{array}
\right)
\]
with a kernel vector
\[
Z=(0,0,0,0,1)^{T}.
\]
As evidently $\Omega_{0}(L_{Z}\Pi_{1})\Omega_{0}=0,$ then $\Pi_{1}$ is
compatible with the pair $(\Pi_{0},\Omega_{0}),$ so the second presymplectic
form is
\begin{align*}
\Omega_{1}  &  =\Omega_{1D}+dH_{1}\wedge dH_{0}\\
&  =\left(
\begin{array}
[c]{ccccc}%
0 & -\frac{1}{2}p_{2} & -q^{1} & -\frac{1}{2}q^{2} & 3(q^{1})^{2}+\frac{1}%
{2}(q^{2})^{2}-c\\
\frac{1}{2}p_{2} & 0 & -\frac{1}{2}q^{2} & 0 & q^{1}q^{2}\\
q^{1} & \frac{1}{2}q^{2} & 0 & 0 & p_{1}\\
\frac{1}{2}q^{2} & 0 & 0 & 0 & p_{2}\\
-3(q^{1})^{2}-\frac{1}{2}(q^{2})^{2}+c & -q^{1}q^{2} & -p_{1} & -p_{2} & 0
\end{array}
\right)  .
\end{align*}
Hence, the bi-presymplectic representation of the Henon-Heiles chain takes the
form%
\[%
\begin{array}
[c]{l}%
\Omega_{0}Y_{0}=0\\
\Omega_{0}Y_{1}=\beta_{1}=\Omega_{1}Y_{0}\\
\Omega_{0}Y_{2}=\beta_{2}=\Omega_{1}Y_{1}\\
\qquad\ \ \ \ \ \ ~0=\Omega_{1}Y_{2}\,\,
\end{array}
\]
where vector fields $Y_{i}$ are
\begin{align*}
Y_{0}  &  =Z=(0,0,0,0,1)^{T}\\
Y_{1}  &  =X_{1}+Z(H_{1})Z=(p_{1},p_{2},-3(q^{1})^{2}-\frac{1}{2}(q^{2}%
)^{2}+c,-q^{1}q^{2},-q^{1})^{T}\\
Y_{2}  &  =X_{1}+Z(H_{2})Z=(\frac{1}{2}q^{2}p_{2},\frac{1}{2}q^{2}p_{1}%
-q^{1}p_{1},\frac{1}{2}p_{2}^{2}-\frac{1}{2}q^{1}(q^{2})^{2},\\
&  -\frac{1}{2}p_{1}p_{2}-\frac{1}{4}(q^{2})^{3}-\frac{1}{2}(q^{1})^{2}%
q^{2}+\frac{1}{2}cq^{2},-\frac{1}{4}(q^{2})^{2})^{T}.
\end{align*}
The chain starts with a kernel vector field $Y_{0}$ of $\Omega_{0}$ and
terminates with a kernel vector field $Y_{2}$ of $\Omega_{1}.$ The restriction
of $\Pi_{0},$ $\Pi_{1d},\Omega_{0}$ and $\Omega_{1}$ to $\mathcal{N}%
=Sp(q^{1},q^{2},p_{1},p_{2})$ are
\[
\pi_{0}=\left(
\begin{array}
[c]{cccc}%
0 & 0 & 1 & 0\\
0 & 0 & 0 & 1\\
-1 & 0 & 0 & 0\\
0 & -1 & 0 & 0
\end{array}
\right)  ,\ \ \ \omega_{0}=\left(
\begin{array}
[c]{cccc}%
0 & 0 & -1 & 0\\
0 & 0 & 0 & -1\\
1 & 0 & 0 & 0\\
0 & 1 & 0 & 0
\end{array}
\right)  ,
\]%
\[
\pi_{1}=\left(
\begin{array}
[c]{cccc}%
0 & 0 & q^{1} & \frac{1}{2}q^{2}\\
0 & 0 & \frac{1}{2}q^{2} & 0\\
-q^{1} & -\frac{1}{2}q^{2} & 0 & \frac{1}{2}p_{2}\\
-\frac{1}{2}q^{2} & 0 & -\frac{1}{2}p_{2} & 0
\end{array}
\right)  ,\ \ \omega_{1}=\left(
\begin{array}
[c]{cccc}%
0 & -\frac{1}{2}p_{2} & -q^{1} & -\frac{1}{2}q^{2}\\
\frac{1}{2}p_{2} & 0 & -\frac{1}{2}q^{2} & 0\\
q^{1} & \frac{1}{2}q^{2} & 0 & 0\\
\frac{1}{2}q^{2} & 0 & 0 & 0
\end{array}
\right)
\]
with the recursion operator $N$ of the form
\[
N=\pi_{1}\omega_{0}=\left(
\begin{array}
[c]{cccc}%
q^{1} & \frac{1}{2}q^{2} & 0 & 0\\
\frac{1}{2}q^{2} & 0 & 0 & 0\\
0 & \frac{1}{2}p_{2} & q^{1} & \frac{1}{2}q^{2}\\
-\frac{1}{2}p_{2} & 0 & \frac{1}{2}q^{2} & 0
\end{array}
\right)
\]
and $N^{\ast}=N^{T}.$

\end{document}